\newcounter{myctr}
\def\myitem{\refstepcounter{myctr}\bibfont\noindent\ifnum\themyctr>9\else\phantom{0}\fi\hangindent17pt\themyctr.\enskip}

\documentclass{ws-ijqi}

\usepackage{graphicx}
\usepackage{amssymb}
\usepackage{bm}
\usepackage{color}
\usepackage[all,cmtip]{xy}

\begin{document}

\markboth{Yu Guo and Heng Fan}
{A generalization of Schmidt number for multipartite states}

\catchline{}{}{}{}{}

\title{A GENERALIZATION OF SCHMIDT NUMBER FOR MULTIPARTITE STATES}

\author{YU GUO}

\address{School of Mathematics and Computer Science, Shanxi Datong University\\ Datong, 037009, China\\
guoyu3@aliyun.com}

\author{HENG FAN}

\address{Institute of Physics, Chinese Academy of Sciences\\ Beijing, 100190, China\\
hfan@iphy.ac.cn}

\maketitle

\begin{history}
\received{Day Month Year}
\revised{Day Month Year}
\end{history}

\begin{abstract}
The Schmidt number is of crucial importance in
characterizing the bipartite pure states. We explore and propose
here a generalization of Schmidt number for states in multipartite
systems. It is shown to be entanglement monotonic and valid for both
pure and mixed states.
In addition, the corresponding generalization of multipartite Schmidt coefficients
is introduced. Our approach is applicable for systems with
arbitrary number of parties and for arbitrary dimensions.
\end{abstract}


\section{Introduction}

Entanglement is shown to play a crucial role in quantum information
processing and quantum computation\cite{Nielsen}. However,
quantifying entanglement is not straightforward, and has become
one of the most significant problems in this
area\cite{Horodecki,Guhne}. Several kinds of entanglement measures
have been proposed for bipartite case\cite{Horodecki,Guhne,Donald}.
Yet there are no operational methods for multipartite states in
general.

In contrast to the bipartite case, the situation is more involved in
the multipartite case. There are many kinds of entanglement. For the
simplest case, a three-qubit state can be either fully separable,
biseparable, or genuinely entangled. An $m$-partite state might have
many different kinds of cases: fully separable, 2-separable,
3-separable, $\dots$, $(m-2)$-separable, and genuinely entangled,
etc. On the other hand, the structure of the local ranke of the
multipartite case is a intricate one. The bipartite pure state
$|\psi\rangle$ is uniquely determined by its reductions but a
tripartite pure state has three single-particle marginals of
inequivalent rank. It is generally difficult to characterize
different types of multipartite entanglement and distinguish them
from each other completely\cite{Gour,Hofmann}.

The \emph{Schmidt number} is indispensable in
characterization and quantification of entanglement associated with
pure
states\cite{Horodecki,Guhne,Donald,Sperling,Buscemi,Sanpera,Fan,Fedorov},
it can be used to
characterize and quantify the degree of bipartite entanglement for
pure state directly\cite{Donald,Sperling}.
In addition, almost any entanglement measure for pure states can be
represented by its Schmidt coefficients\cite{Donald}, such as entanglement formation, concurrence, etc.
Then a natural idea in mind is that whether there is a corresponding
quantity for the multipartite states.
Unfortunately,
the Schmidt decomposition is not valid for multipartite case. Only
rare pure states in the multipartite case admit the
Schmidt decomposition form\cite{Peres1995,Thapliyal}
\begin{eqnarray}
|\psi\rangle=\sum_{k=1}^{r}
\lambda_k|e^{(1)}_k\rangle|e^{(2)}_k\rangle\otimes\cdots\otimes|e^{(m)}_k\rangle,
\end{eqnarray}
where $r\leq \min\{N_1,N_2,\dots,N_m\}$, $N_i$ denotes the dimension of the
$i-$th subsystem, $\{|e_k^{(i)}\rangle\}$ is an orthonormal set of the $i-$th state space.
For the simplest three-partite case, any
three-qubit pure state can be written as\cite{Acin} $|\psi\rangle
=\lambda_0|000\rangle+\lambda_1e^{i\theta}|100\rangle
+\lambda_2|101\rangle+\lambda_3|110\rangle +\lambda_4|111\rangle$,
where $\lambda_i\geq0$, $\sum_i\lambda_i^2=1$,
$\theta\in[0,\pi]$, which is not a form of Schmidt
decomposition. That is, there is no correspondence of Schmidt number
for multipartite case in general.
In Ref.~\refcite{Eisert}, a generalized Schmidt number for a multipartite state $|\psi\rangle$
is defined to be the minimal necessary number of
summands in the representation of it as a sum of separable
states. It is shown to be a multipartite entanglement measure but the scenario is not based on the Schmidt decomposition.
In this paper, we try to extend the
Schmidt number to multipartite systems according to the Schmidt decomposition of the subsystems.
And as a closely related concept, the corresponding multipartite Schmidt coefficients will be discussed.
As desired, we show that the generalized Schmidt number can be used to quantify entanglement for multipartite states as well.
That is, we propose a new entanglement measure for multipartite states.

The rest of this paper is organized as follows. In Sec. 2,
we review the origin Schmidt number and the Schmidt coefficients for the bipartite case.
In Sec. 3, we give our method of extending the Schmidt number to the multipartite systems.
We begin with the tripartite case and then discuss the generalized case.
Both the pure states and the mixed states are considered. We show that our generalized Schmidt number is an entanglement monotone.
Then, in Sec. 4, we construct the multipartite Schmidt coefficients based on our scenario of the multipartite Schmidt number.
Sec. 5 lists two kinds of examples: the W state and the GHZ state. Finally, in Sec. 6,
we draw the conclusion.

\if
Almost any entanglement measure for pure states can be
represented by its Schmidt coefficients\cite{Donald}. The von
Neumann entanglement entropy of bipartite pure state $|\psi\rangle$
is
\begin{eqnarray*}
E(|\psi\rangle)=S(\rho_1)
=S(\rho_2)=-\sum_k\lambda_k^2\log_2\lambda_k^2,
\end{eqnarray*}
where $S(\cdot)$ denotes the von Neumann entropy.
The entanglement of formation for mixed
states $\rho$ is defined as
\begin{eqnarray*}
E_f(\rho)
=\inf\limits_{\mathcal{D}_2(\rho)}\sum\limits_kp_kE(|\psi_k\rangle),
\end{eqnarray*}
where $\inf\limits_{\mathcal{D}_2(\rho)}$
denotes the infimum over all pure ensembles of  $\rho$.
\fi

\section{Preliminary}

Throughout this paper, we only consider the finite-dimensional case since the Schmidt number for the continuous-variable system may be $\infty$.
Let $|\psi\rangle=\sum_{i,j}d_{ij}|i_1\rangle|j_2\rangle$ be a pure
state lives in $\mathbb{C}^{N_1}\otimes\mathbb{C}^{N_2}$ and
$|\psi\rangle=\sum_{k}\lambda_{k}|e_k\rangle|f_k\rangle$ be its
Schmidt decomposition\cite{Schmidt}, where $\{|i_1\rangle\}$ and $\{|j_2\rangle\}$
are the standard computational bases of $\mathbb{C}^{N_1}$ and
$\mathbb{C}^{N_2}$ respectively, $\{|e_k\rangle\}$ and $\{|f_k\rangle\}$
are orthonormal sets of $\mathbb{C}^{N_1}$ and
$\mathbb{C}^{N_2}$ respectively. Then the Schmidt number of
$|\psi\rangle$ is defined by
\begin{eqnarray}
R_\psi={\rm rank}(\rho_1)={\rm rank}(\rho_2),
\end{eqnarray}
where $\rho_i$ denotes the reduced state of the $i-$th part.
$\lambda_k$s are called the Schmidt
coefficients of $|\psi\rangle$.
It is clear that $R_\psi$ coincides with the rank of the
coefficient matrix of $|\psi\rangle$,
i.e., $R_\psi={\rm rank}(D)$, $D=[d_{ij}]$;
$R_\psi$ is also the length of the Schmidt
decomposition of $|\psi\rangle$. For mixed state $\rho$ acting on $\mathbb{C}^{N_1}\otimes\mathbb{C}^{N_2}$, the Schmidt number is defined by\cite{Horodecki2000pra}
\begin{eqnarray}
R_\rho=\inf\limits_{\mathcal{D}_2(\rho)}\max\limits_{\psi_i}R_{\psi_i},
\end{eqnarray}
where $\mathcal{D}_2(\rho)=\{p_i,|\psi_i\rangle: \rho
=\sum_ip_i|\psi_i\rangle\langle\psi_i|\}$ is for all pure ensembles
of the bipartite state $\rho$.
It is shown that $R_\rho$ is entanglement monotonic\cite{Horodecki2000pra}.

\section{The Schmidt number for multipartite case}

\begin{table}[ph]
\tbl{\label{tab:2}The Schmidt number of the
three qubit pure state ($r_i$ denotes the rank
of $\rho_i$, $`i-j$' means $\rho_{ij}$ is not a
pure separable state, $`i~~~j-k$' means
$\rho_{jk}=|\psi_{jk}\rangle\langle\psi_{jk}|$ is an
entangled pure state, etc.)}
{\begin{tabular}{@{}cccc@{}} \toprule
Type & Model            &  Reductions                                         & $R_\psi$\\
\colrule
$1|2|3$      &$1~~~ 2~~~3$      & $r_i=1$                                             & 1 \\
$1|23$       &$1~~~2-3$         & $r_1=1$, $r_2=r_3=2$                                & 2\\
$12|3$       &$1-2~~~3$         & $r_3=1$, $r_1=r_2=2$                                & 2\\
$2|13$       &$2~~~1-3$         & $r_2=1$, $r_1=r_3=2$                                & 2\\
GE           &figure (a) & $r_i=2$, $\rho_{\bar{i}}$ is separable, $i=1,2,3$   & 3\\
GE           &figure (a) & $r_i=2$, $\rho_{\bar{i}}$ is entangled for some $i$ & 4\\
\botrule
\end{tabular}}
\end{table}
We develop a method of extending the Schmidt number to multipartite
case. We consider the three qubit case first. If $|\psi\rangle$ is
fully separable (i.e., $1|2|3$ separable), then
$|\psi\rangle=|\psi_1\rangle|\psi_2\rangle|\psi_3\rangle$. In such a
case, the Schmidt number of $|\psi\rangle$ can be defined to be 1.
Suppose $|\psi\rangle$ is bi-separable, without loss of generality,
we assume that it is $1|23$ separable and not fully separable, i.e.,
$|\psi\rangle=|\psi_1\rangle|\psi_{23}\rangle$. Since
$|\psi_{23}\rangle$ is a bipartite entangled pure state, it has
Schmidt number $R_{\psi_{23}}=2$, and then $R_\psi$ can be viewed as
$R_{\psi_{23}}$, namely, $R_\psi=2$. If $|\psi\rangle$ is genuinely
entangled, then ${\rm rank}(\rho_1)={\rm rank}(\rho_{23}) ={\rm
rank}(\rho_2)={\rm rank}(\rho_{13})= {\rm rank}(\rho_3)={\rm
rank}(\rho_{12})=2$, where $\rho_{\gamma}$ denotes the reduction of
the $\gamma-$subsystem(s). If $\rho_{12}$, $\rho_{23}$ and
$\rho_{13}$ are separable, then
$R_{\rho_{12}}=R_{\rho_{13}}=R_{\rho_{23}}=1$, we thus view $R_\psi$
as ${\rm rank}(\rho_1)+R_{\rho_{23}}= {\rm
rank}(\rho_2)+R_{\rho_{13}} ={\rm rank}(\rho_3)+R_{\rho_{12}}=3$. If
$\rho_{23}$, or $\rho_{12}$, or $\rho_{13}$ is entangled, then
$\max\{R_{\rho_{12}},R_{\rho_{23}},R_{\rho_{13}}\}=2$, we thus view
$R_\psi$ as ${\rm rank}(\rho_1)+2=4$. That is, there are four types
of entanglement indeed for the three qubit case (see
Table~\ref{tab:2}). (For $i\in\{1,2,3,\dots,m\}$, we denote by
$\bar{i}$ the combination consisting of all elements in
$\{1,2,\dots,m\}-\{i\}$, for instance, if $m=4$, $i=(2)$, then
$\bar{i}=(134)$.)

\begin{table}[ph]
\tbl{\label{tab:3}The Schmidt number of $|\psi\rangle$ in
the three-partite system.}
{\begin{tabular}{@{}ccc@{}} \toprule
Type    &  Reductions                      & $R_\psi$\\
\colrule
$1|2|3$         & $r_i=1$                          & 1\\
$1|23$          & $r_1=1$, $r_2=r_3=R_{\psi_{23}}$ & $R_{\psi_{23}}$\\
$12|3$          & $r_3=1$, $r_1=r_2=R_{\psi_{12}}$ & $R_{\psi_{12}}$\\
$2|13$          & $r_2=1$, $r_1=r_3=R_{\psi_{13}}$ & $R_{\psi_{13}}$\\
GE              & $r_i\geq 2$, $i=1$,2,3               & $\max\limits_i(r_i+R_{\rho_{\bar{i}}})$\\
\botrule
\end{tabular}}
\end{table}

\begin{table}[ph]
\tbl{\label{tab:4}The Schmidt number of
$|\psi\rangle$ in the four-partite system. }
{\begin{tabular}{@{}cccc@{}} \toprule
Type  &Model          &  Local rank   & $R_\psi$\\
\colrule
$1|2|3|4$   & $1~~~2~~~3~~~4$ & $r_i=1$       & 1\\
$1|23|4$    & $1~~~2-3~~~4$   &$r_1=r_4=1$    & $R_{\psi_{23}}$\\
$12|3|4$    & $1-2~~~3~~~4$   &$r_3=r_4=1$    & $R_{\psi_{12}}$\\
$2|3|14$    & $2~~~3~~~1-4$   &$r_2=r_3=1$    & $R_{\psi_{14}}$\\
$2|13|4$    & $2~~~1-3~~~4$   &$r_2=r_4=1$    & $R_{\psi_{13}}$\\
$1|3|24$    & $1~~~3~~~2-4$   &$r_1=r_3=1$    & $R_{\psi_{24}}$\\
$1|2|34$    & $1~~~2~~~3-4$   &$r_1=r_2=1$    & $R_{\psi_{34}}$\\
$1|234$     & $1~~~2-3-4$     & $r_1=1$       & $R_{\psi_{234}}$\\
$2|134$     & $2~~~1-3-4$     &$r_2=1$        & $R_{\psi_{134}}$\\
$3|124$     & $3~~~1-2-4$     &$r_3=1$        & $R_{\psi_{124}}$\\
$4|123$     & $4~~~1-2-3$     &$r_4=1$        & $R_{\psi_{123}}$\\
$12|34$     & $1-2~~~3-4$     &$r_i\geq2$     & $R_{\psi_{12}}+R_{\psi_{34}}$\\
$13|24$     & $1-3~~~2-4$     &$r_i\geq2$     & $R_{\psi_{13}}+R_{\psi_{24}}$\\
$14|23$     & $1-4~~~2-3$     &$r_i\geq2$     & $R_{\psi_{14}}+R_{\psi_{23}}$\\
GE          & figure (b) &$r_i\geq2$ & $\max\limits_i(r_i+R_{\rho_{\bar{i}}})$\\
\botrule
\end{tabular}}
\end{table}

We now move to the $N_1\otimes N_2\otimes N_3$ case. We may assume
that $N_1\leq N_2\leq N_3$. If $|\psi\rangle$ is fully separable,
then $|\psi\rangle=|\psi_1\rangle|\psi_2\rangle|\psi_3\rangle$ and
thus the Schmidt number of $|\psi\rangle$ can be considered to be 1.
If $|\psi\rangle$ is bi-separable, without loss of generality, we
assume that it is $1|23$ separable and not fully separable, i.e.,
$|\psi\rangle=|\psi_1\rangle|\psi_{23}\rangle$. Since
$|\psi_{23}\rangle$ is a bipartite entangled pure state, we let
$R_{\psi_{23}}=t$, $2\leq t\leq N_2$, then $R_\psi$ can be viewed as
$R_{\psi_{23}}$, that is, $R_\psi=t$. If $|\psi\rangle$ is genuinely
entangled, then ${\rm rank}(\rho_1)={\rm rank}(\rho_{23})=i\geq 2$,
${\rm rank}(\rho_2)={\rm rank}(\rho_{13})=j\geq2$ and ${\rm
rank}(\rho_3)={\rm rank}(\rho_{12})=k\geq2$. If $\rho_{12}$,
$\rho_{23}$ and $\rho_{13}$ are separable, then
$R_{\rho_{12}}=R_{\rho_{13}}=R_{\rho_{23}}=1$, we thus view $R_\psi$
as $\max_i({\rm rank}(\rho_i)+R_{\rho_{\bar{i}}})\geq3$. If
$\rho_{23}$, or $\rho_{12}$, or $\rho_{13}$ is entangled, then
$\max\{R_{\rho_{12}},R_{\rho_{23}},R_{\rho_{13}}\}\geq2$, we thus
view $R_\psi$ as $\max_i({\rm
rank}(\rho_i)+R_{\rho_{\bar{i}}})\geq4$. That is, there are at most
$N_1+N_3$ types of entanglement for the $N_1\otimes N_2\otimes N_3$ case (see
Table~\ref{tab:3}).


\noindent{\bf Note.}  If $m>3$, then there exist $|\psi\rangle$ and
$|\phi\rangle$ in $m-$partite systems, such that $|\psi\rangle$ is
$k-$separable, $|\phi\rangle$ is genuinely entangled but
$R_\psi>R_\phi$.


A natural way of generalizing the Schmidt number to mixed states
$\rho$ acting on
$\mathbb{C}^{N_1}\otimes\mathbb{C}^{N_2}\otimes\mathbb{C}^{N_3}$ can
be defined by the convex roof structure
\begin{eqnarray}
R_\rho:
=\inf\limits_{\mathcal{D}_3(\rho)}\max\limits_{\psi_i}R_{\psi_i},\label{x}
\end{eqnarray}
where the minimization is over all ensemble of $\rho$
(Hereafter, we denote by $\mathcal{D}_k(\rho)$ the set of all ensembles of the $k$-partite state $\rho$).
It is reasonable as this means that
$\rho$ cannot be obtained by mixing pure states with
Schmidt number lower than $R_{\rho}$ and
that there exists an ensemble with Schmidt number at most $R_\rho$
to reach the state.
It is clear that $R_\rho$ is an entanglement
monotone since local rank of pure state is non-increasing under local
operations and classical communication (LOCC)\cite{Lo}.

Now we can establish the Schmidt number for the four-partite system
as Table~\ref{tab:4}. Analogously, we can define the Schmidt number
for mixed states via the convex roof structure as Eq.~(\ref{x}).
This approach can be extended to $m-$partite case step by step for
both pure and mixed states. That is, for
$|\psi\rangle\in\mathbb{C}^{N_1}\otimes\mathbb{C}^{N_2}\otimes\cdots\otimes\mathbb{C}^{N_m}$,
the $R_{\psi}$ can be defined as the program discussed above:
if $|\psi\rangle$ is not genuinely entangled and assume with on loss of generality that it is $12|34|5\cdots m$ separable (resp. $12|3|4|5\cdots m$ separable),
then $R_{\psi}=R_{\psi_{12}}+R_{\psi_{34}}+R_{\psi_{5\cdots m}}$ (resp. $R_{\psi}=R_{\psi_{12}}+R_{\psi_{5\cdots m}}$); if $|\psi\rangle$ is genuinely entangled, then
$R_{\psi}=\max\limits_{i}(r_i+R_{\rho_{\bar{i}}})$. For
mixed state $\rho$ acting on
$\mathbb{C}^{N_1}\otimes\mathbb{C}^{N_2}\otimes\cdots\otimes\mathbb{C}^{N_m}$,
it can be defined by the convex roof structure
\begin{eqnarray}
R_\rho:
=\inf\limits_{\mathcal{D}_m(\rho)}\max\limits_{\psi_i}R_{\psi_i},\label{a}
\end{eqnarray}
where the minimization is over all ensemble of $\rho$.
We now can conclude the following result.

\smallskip

\noindent{\bf Theorem:} The Schmidt number defined as Eq.~(\ref{a}) is an
entanglement monotone.

\smallskip

Similar to the bipartite case, a multipartite $\rho$ is fully
separable iff $R_\rho=1$. We have now established a complete
hierarchy of Schmidt numbers that quantify the dimensions of the
entanglement. The bipartite Schmidt number can be viewed as an
entanglement measure since the Schmidt number fully
reflects the dimensional of entanglement\cite{Sperling}. From this
point of view, the generalized Schmidt number can also be viewed as
an entanglement measure for the multipartite case.

We illustrate the `structure' of the Schmidt number for genuinely entanglement with the
following figures. $`r_i-R_{ij}-r_j$' means $\rho_{ij}$ is a mixed state
with Schmidt number $R_{ij}$. For the tripartite case, the Schmidt number
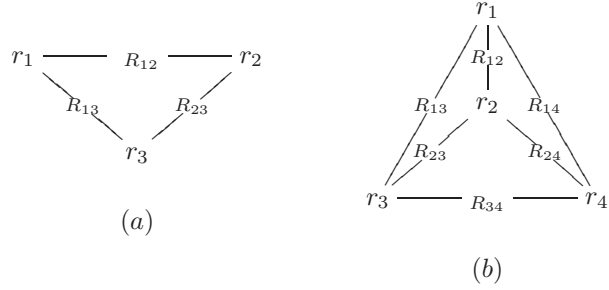
\begin{eqnarray*}
&\begin{array}{c}\xymatrix{
r_1 \ar@{-}[rd]|-{R_{{13}}}  \ar@{-}[r]
&{~_{R_{{12}}}} \ar@{-}[r]&
r_2 \ar@{-}[l]\ar@{-}[ld]|-{R_{{23}}}\\
&r_3&}\\ \\
(a)\end{array}\quad\quad\quad
\begin{array}{c}\xymatrix{
&r_1 \ar@{-}[ldd]|-{R_{13}}  \ar@{-}[d]|-{R_{{12}}} \ar@{-}[rdd]|-{R_{14}}\\
&r_2 \ar@{-}[ld]|-{R_{23}} \ar@{-}[rd]|-{R_{24}}\\
r_3\ar@{-}[r] &_{R_{34}}&\ar@{-}[l]r_4 }\\ \\
(b)\end{array} &
\end{eqnarray*}
can be `explained' by Figure (a): $R_\psi=\max\limits_{i}\{r_i+R_{\bar{i}}\}$. For the
four-partite pure state $|\psi\rangle$,
$R_\psi$ is determined by $r_i$
and $R_{\bar{i}}$ while
$R_{\bar{i}}$ is decided by the
Schmidt number of the six bipartite
reductions and the four single
part reductions. It is clear that
$R_\psi\leq\max\limits_{i\neq j}\{r_i+r_j+R_{\bar{ij}}\}$
(see the `relation' among $r_i$, $r_j$ and $R_{\bar{ij}}$ in Figure (b)).

\smallskip

\noindent{\bf Note.}  It is straightforward that the generalized Schmidt
number is invariant under the invertible SLOCC (stochastic local
operations and classical communication) since invertible SLOCC
preserves the rank of the reduction\cite{Dur} (also see in Refs.
\refcite{Cornelio,Lidafa,Wangshuhao}).

\section{The Schmidt coefficients for multipartite case}

For the bipartite system, almost any entanglement measure or even any quantum correlation for pure states can be
represented by its Schmidt coefficients\cite{Donald,Luo,Wu2014,Guowu2014,Guo2014ijtp}. In this section, we discuss the Schmidt coefficients
for the multipartite case
based on the scenario of the multipartite Schmidt number. We begin
with the tripartite case. Let $|\psi\rangle$ be a pure state in a
$N_1\otimes N_2\otimes N_3$ system. If it is fully separable, the
Schmidt coefficient is $\{1\}$. If it is $1|23$ separable, the
Schmidt coefficients are defined as that of $|\psi_{23}\rangle$;
similarly, we can define it for the types $12|3$ and $2|13$. If it
is genuinely entangled, we assume without loss of generality that
\begin{eqnarray*}
R_\psi={\rm rank}(\rho_{t_1})+R_{\rho_{\bar{t_1}}}
={\rm rank}(\rho_{t_2})+R_{\rho_{\bar{t_2}}}
=\cdots
={\rm rank}(\rho_{t_k})+R_{\rho_{\bar{t_k}}}.
\end{eqnarray*}
Let $\tilde{\rho}_{t_i}
=\frac{1}{\sqrt{2}}\rho_{t_i}^{\frac{1}{2}}$, and let
\begin{eqnarray*}
\sigma(\tilde{\rho}_{t_i})
=\{\lambda_p^{(i)}\}_{p=1}^{{\rm rank}(\tilde{\rho}_{t_i})},
\end{eqnarray*}
where $\sigma(\cdot)$ denotes the set of
eigenvalues of the described matrix.
Let $|\psi_1^{(t_i)}\rangle$, $|\psi_2^{(t_i)}\rangle$,
$\dots$, $|\psi_r^{(t_i)}\rangle$ be elements
in the pure state ensembles of
$\rho_{\bar{t_i}}$ such that
$R_{\psi_j^{(t_i)}}=R_{\rho_{\bar{t_i}}}$, $j=1$, 2, $\dots$, $r$.
Assume that
\begin{eqnarray*}
E(|\psi_{j_0}^{(t_i)}\rangle)=\max\limits_jE(|\psi_{j}^{(t_i)}\rangle),
\end{eqnarray*}
$\mathcal{S}_C(|\psi_{j_0}^{(i)}\rangle)=\{\lambda_q^{(i)}\}$
and $\delta_q^{(i)}=\frac{\lambda_q^{(i)}}{\sqrt{2}}$,
where $\mathcal{S}_C(x)$ denotes
the Schmidt coefficients of $x$.
We write
\begin{eqnarray*}
E_i(|\psi\rangle):=
\left\{\begin{array}{ll}
-\sum\limits_p(\lambda_p^{(i)})^2\log_2(\lambda_p^{(i)})^2
-\sum\limits_q(\delta_q^{(i)})^2\log_2(\delta_q^{(i)})^2 & {\rm if}\ R_{\rho_{\bar{t_i}}}>1,\\
-\sum\limits_p(\lambda_p^{(i)})^2\log_2(\lambda_p^{(i)})^2
+\frac{1}{2} &{\rm if}\ R_{\rho_{\bar{t_i}}}=1.
\end{array}\right.
\end{eqnarray*}
If $E_i$ reaches a maximum for some $i$ with $R_{\rho_{\bar{t_i}}}=1$,
\begin{eqnarray}
\sigma(\tilde{\rho}_{t_i})\bigcup \{\frac{1}{\sqrt{2}}\}
\end{eqnarray}
is defined to be the Schmidt coefficients of $|\psi\rangle$;
If $E_i$ reaches a maximum for some $i$ with $R_{\rho_{\bar{t_i}}}>1$,
denote $\mathcal{S}_C(|\psi_{j_0}^{(i)}\rangle)$ by $\mathcal{S}_C^{t_i}$,
\begin{eqnarray}
\sigma(\tilde{\rho}_{t_i})\bigcup \mathcal{S}_C^{t_i}
\end{eqnarray}
is defined to be the Schmidt coefficients of $|\psi\rangle$.
The ratio `$\frac{1}{\sqrt{2}}$' here guarantees that the
Schmidt coefficients are normalized, i.e.,
the sum of the squares is 1.

For the four-partite system, if it is not genuinely entangled, then
it reduces to the three-partite case. For example, if $|\psi\rangle$
is $1|234$ separable, then
$|\psi\rangle=|\psi_1\rangle|\psi_{234}\rangle$. So we can define
the Schmidt coefficients as that of $|\psi_{234}\rangle$. If
$|\psi\rangle$ is $12|34$ separable, then
$|\psi\rangle=|\psi_{12}\rangle|\psi_{34}\rangle$. In such a case,
we define the Schmidt coefficients to be
$\sigma(\tilde{\rho}_{1})\bigcup\sigma(\tilde{\rho}_{3})$. In these
cases, the von Neumann entanglement entropy is clear. If it is
genuinely entangled, we let $|\phi_0^{(i)}\rangle$ be an element in
the pure state ensembles of $\rho_{\bar{i}}$ such that $R_\psi={\rm
rank}(\rho_i)+R_{\rho_{\bar{i}}}$,
$R_{\rho_{\bar{i}}}=R_{\phi_0^{(i)}}$ and $E(|\phi_0^{(i)}\rangle)$
reaches the maximal over all elements $|\phi^{(i)}\rangle$s in the
ensembles of $\rho_{\bar{i}}$ such that
$R_{\rho_{\bar{i}}}=R_{\phi^{(i)}}$, where $E(|\phi_0^{(i)}\rangle)$
is defined as $E(|\psi_0^{(i)}\rangle):=\max\limits_j
E_j(|\psi_0^{(i)}\rangle)$, $1\leq j\leq 4$. If
\begin{eqnarray}
S(\tilde{\rho}_{i_0})+E(|\phi_0^{(i_0)}\rangle)
=\max\limits_i\{S(\tilde{\rho}_{i})+E(|\phi_0^{(i)}\rangle)\},
\end{eqnarray}
we let $\mathcal{S}_C(|\phi_0^{(i_0)}\rangle)=\{\gamma_k\}$
and $\tilde{\mathcal{S}}_C^{i_0}=\{\tilde{\gamma}_k\}$ with
$\tilde{\gamma}_k=\frac{\gamma_k}{\sqrt{2}}$.
Then we call
\begin{eqnarray}
\sigma(\tilde{\rho}_{i_0})\bigcup \tilde{\mathcal{S}}_C^{i_0} \label{8}
\end{eqnarray}
the Schmidt coefficients of $|\psi\rangle$.

\smallskip

\noindent{\bf Note.}  (i) $\rho$ is fully separable iff the Schmidt
coefficients is $\{1\}$; (ii) the number of the Schmidt coefficients
coincides with the Schmidt number; (iii) the Schmidt coefficients may
not be unique.

\section{Examples}

We end our discussion with some examples. Two well known three qubit
states are the W state and the GHZ state,
\begin{eqnarray*}
|W_3\rangle&=&\frac{1}{\sqrt{3}}(|001\rangle+|010\rangle+|100\rangle),\\
|{\rm GHZ_3}\rangle&=&\frac{1}{\sqrt{2}}(|000\rangle+|111\rangle).
\end{eqnarray*}
For the state $|W\rangle$, one can easily see that
any coefficient matrix of any
bipartite splitting is not of rank-one,
so it is genuinely entangled.
From Table~\ref{tab:2}, $R_{W_3}=4$.
The state $|{\rm GHZ}_3\rangle$ is
also genuinely entangled.
Table~\ref{tab:2} indicates
$R_{{\rm GHZ}_3}=3$.
In Ref.~\refcite{Dur}, it is proved
that $|W_3\rangle$ and $|{\rm GHZ}_3\rangle$
are two types of genuinely entangled states
under SLOCC classification, which meets our results.
The Schmidt coefficients of $|W_3\rangle$ are
$\{\frac{1}{\sqrt{3}},\frac{1}{\sqrt{6}},
\frac{1}{2},\frac{1}{2}\}$.
The Schmidt coefficients of $|{\rm GHZ}_3\rangle$ are
$\{\frac{1}{2},\frac{1}{2},\frac{1}{\sqrt{2}}\}$.
In addition, the entanglement dimensionality vector\cite{Huber}
of $|W_3\rangle$, $(2$, 2, $2)$,
coincides with that of $|{\rm GHZ}_3\rangle$.
From this point of view the generalized Schmidt number provides
a more strict classification of multipartite
states than the scenario of entanglement dimensionality vector
proposed in Ref.~\refcite{Huber}.
In addition, it is worth noticing that
the Schmidt number is different from the \emph{collectibility} proposed in Ref.~\refcite{Rudnicki}
since the collectibility of $|{\rm GHZ}_3\rangle$ is larger than that of $|W_3\rangle$.

For the $m-$qubit W-state $|W_m\rangle$ and the GHZ state
\begin{eqnarray*}
|W_m\rangle&=&\frac{1}{\sqrt{m}}
(|0\cdots01\rangle+|0\cdots010\rangle+|1\cdots00\rangle),\\
|{\rm GHZ}_m\rangle&=&\frac{1}{\sqrt{2}}
(|0\rangle^{\otimes m}+|1\rangle^{\otimes m}),
\end{eqnarray*}
one can check that
$R_{W_m}=2(m-1)$
and $R_{{\rm GHZ}_m}=3$.
It is worth mentioning here that
all the reductions of the $|W_m\rangle$ are
genuinely entangled while all the reductions of
the $|{\rm GHZ}_m\rangle$ are (fully) separable.
Similarly, for the generalized GHZ
state in the $d^{\otimes m}$ system,
the Schmidt number of
$|{\rm GHZ}_m^{(d)}\rangle
=\frac{1}{\sqrt{d}}\sum_{i=1}^d(|i\rangle^{\otimes m})$
is $d+1$.

\section{Conclusion}

In this paper, the generalizations of the Schmidt number and the Schmidt coefficients for
multipartite case are established from a mathematical point-of-view. We showed that
the generalized Schmidt number is a well-defined entanglement measure since it
is entanglement monotonic. Our results may shed new lights on the
task of multipartite systems: the multipartite states can be
classified via the generalized Schmidt number.
We also hope that our mathematic scenario of the Schmidt number
may induce some exact physical or operational meaning.

Going further, one can define the generalized entanglement formation in
terms of the Schmidt coefficients. That is, if the Schmidt
coefficients of $|\psi\rangle$ are $\{\eta_i\}$, then we can define
the generalized entanglement of formation by
$E(|\psi\rangle):=-\sum_i\eta_i^2\log_2\eta_i^2$. It can be extended
to mixed states via the convex roof structure. (Note that although
the Schmidt coefficients may not be unique the generalized von
Neumann entanglement entropy is unique.) The origin entanglement of
formation for the bipartite case is an entanglement monotone, we
conjecture that the generalized entanglement of formation is an
entanglement monotone (the proof maybe a hard work due to the
complex structure of both the multipartite states and the
multipartite LOCC).

\section*{Acknowledgments}

Y. Guo is supported by the National Natural Science Foundation of China under Grants No. 11301312 and 11171249,
the Natural Science Foundation of Shanxi
under Grant No. 2013021001-1 and 2012011001-2, and the Research start-up fund for Doctors of Shanxi Datong University
under Grant No. 2011-B-01.
H. Fan is
supported by the `973' program (Grant No. 2010CB922904).




\end{document}